\documentclass[twocolumn,aps,prl,showpacs,preprintnumbers,floatfix]{revtex4}

\usepackage[latin9]{inputenc}
\usepackage{amsmath}
\usepackage{amssymb}
\usepackage{graphicx}

\usepackage{xspace}
\newcommand{\eg}{{e.g.,\/}\xspace}
\newcommand{\ie}{{i.e.,\/}\xspace}

\newcommand{\eq}[1]{(\ref{#1})}
\newcommand{\Eq}[1]{Eq.~(\ref{#1})}
\newcommand{\Eqs}[1]{Eqs.~(\ref{#1})}
\newcommand{\Fig}[1]{Fig.~\ref{#1}}

\newcommand{\Ref}[1]{Ref.~\cite{#1}}

\newcommand{\pd}{\partial}
\newcommand{\del}{\vec{\nabla}}
\newcommand{\mc}[1]{\mathcal{#1}}
\newcommand{\oper}[1]{\hat{\vec{#1}}}
\renewcommand{\vec}[1]{{\boldsymbol{\rm #1}}}
\newcommand{\msection}[1]{\textit{#1.}\ ---\ }

\begin{document}

\title{Ladder Climbing and Autoresonant Acceleration of Plasma Waves}

\author{I. Barth$^a$, I. Y. Dodin$^{a,b}$, and N. J. Fisch$^{a,b}$}

\affiliation{
$^a$Princeton Plasma Physics Laboratory, Princeton University, Princeton, New Jersey 08543, USA \\
$^b$Department of Astrophysical Sciences, Princeton University, Princeton, New Jersey 08540, USA
}

\date{\today}

\begin{abstract}

When the background density in a bounded plasma is modulated in time, discrete modes become coupled.    
Interestingly, for appropriately chosen modulations, the average plasmon energy might be made to grow in a ladder-like manner, achieving up-conversion or down-conversion of the plasmon energy. This reversible process is identified as a classical analog of the effect known as quantum ladder climbing, so that the efficiency and the rate of this process can be written immediately by analogy to a quantum particle in a box. In the limit of densely spaced spectrum, ladder climbing transforms into continuous autoresonance; plasmons may then be manipulated by chirped background modulations much like electrons are autoresonantly manipulated by chirped fields. By formulating the wave dynamics within a universal Lagrangian framework, similar ladder climbing and autoresonance effects are predicted to be achievable with general  linear waves in both plasma and other media. 

\end{abstract}


\maketitle

\msection{Introduction} Quantum mechanics is well known to be closely related to the mechanics of classical waves \cite{Dragoman,Tracy,ilya_PLA14}. This permits applying common techniques for manipulating quantum and classical systems and helps bridging seemingly different areas of physics. One important technique to study in this context is ladder climbing (LC), which is the successive transfer of quanta through nonequally-spaced energy levels due to an oscillating driving force with chirped frequency \cite{Marcus_th,ido5}. The system energy changes with time in a ladder-like manner during LC, with each transition described by the famous Landau-Zener (LZ) theory \cite{LZ}. In the limit of continuous spectra, the effect has been widely known as classical autoresonance (AR), enjoying numerous applications in physics of plasmas \cite{Deutsch_lazar101_ALPHA_Baker,lazar111_ido2}, fluids \cite{Ben-David}, Josephson junctions \cite{lazar121_ido4}, optics \cite{Barak}, and even planetary dynamics \cite{Malhotra}. In contrast, the discrete nature of LC is visible only in systems with sufficiently discrete spectra and, so far, has been studied exclusively in quantum contexts \cite{Marcus_exp_ido6,Marcus_th,ido5,ido7,ido8}. Whether classical systems can exhibit LC has remained an open question.

Here we report the first theoretical prediction of LC in a classical system, namely, in an ensemble of plasma waves. For simplicity, we consider one-dimensional collisionless plasma, with nondissipative Langmuir waves, whose spectrum is quantized due to the boundary conditions. We derive a Schr\"{o}dinger-type equation for the ``plasmon wave function'', which is a classical measure of the electric field whose norm (the total wave action) is manifestly conserved under mode coupling. 
This system is mathematically equivalent to a quantum particle in electrostatic potential. Hence, plasmons can be manipulated by resonant modulation of the underlying medium, much like electrons and molecules are manipulated by resonant external fields \cite{Breizman_Schmit,shneider}. In particular, we show that plasmons can exhibit both LC and AR and can be controllably transported up and down in momentum space. Finally, we report a unifying Lagrangian formulation of the problem that paves the way for applying these techniques to general classical waves.

\msection{Basic equations} For simplicity, we consider an electron plasma described by a hydrodynamic model. The equations for the electron density $n_e$, electron flow velocity $u_e$, and the electric field $E$ are then as follows:
\begin{eqnarray} 
	\frac{\partial n_e}{\partial t}+\frac{\partial }{\partial x}(n_e u_e) & = & 0, \label{dn_dt}\\
	\frac{\partial}{\partial t}(n_e u_e)+\frac{\partial }{\partial x}(n_e u_e^2)
	& = &-\frac{e}{m_e}n_e E -\frac{1}{m_e}\frac{\partial P}{\partial x}, \label{dnu_dt}\\
	\frac{\partial E}{\partial x}&=&-4\pi e (n_e-Zn_i). \label{dE_dx}
\end {eqnarray}
Here, $-e$ and $m_e$ are the electron charge and mass, $P$ is the electron pressure, $Ze$ is the ion charge, and $n_i$ is the ion density. We neglect high-frequency oscillations of $n_i$ and consider $n_i$ to be a slow function, $Z n_i=n_0+n_d(t,x)$; here $n_0$ is the unperturbed electron density, and $n_d$ is a prescribed driving modulation. Such a modulation can be created by external fields, \eg by means of ponderomotive forces. Then, $n_e=n_0+n_d(t,x)+n(t,x)$, where $n \ll n_0$ determines a small uncompensated charge density due to electron inertia. For simplicity, we adopt an isentropic model, $P = P(n_e)\approx P_0+3 m_e v_{th}^2 (n+n_d)+R(n+n_d)^2/2$; here $v_{th}$ is the electron thermal speed, and $R=\partial^2 P/\partial n^2 |_{n_0}$, which are considered constant. We assume hard-wall boundary conditions, so $u_e|_{x=0}=u_e|_{x=\ell}=0$, where $\ell$ is the plasma length. We also assume $E|_{x=0}=E|_{x=\ell}=0$, so any field is representable as 
a series of $\sin(k_m x)$, where $k_m = \pi m/\ell$. Then, according to \Eqs{dn_dt}-\eq{dE_dx}, the boundary conditions for the density must be $\partial_x n |_{x=0}=\partial_x n |_{x=\ell}=0$, so $n$ is a series of $\cos(k_m x)$. We consider the external driving modulation to be the $N$th standing-wave mode,
\begin{gather}\label{n_d}
 n_d(x,t)=n_0 A\cos\left(k_N x\right)\cos\varphi_d(t).
\end{gather}
We assume that $A\ll1$ and $\omega_d=\dot{\varphi}\ll\omega_p$, where $\omega_p^2 = 4\pi e^2 n_0 /m_e$. Hence, \Eqs{dn_dt}-\eq{dE_dx} can be combined in a single equation. Specifically, let us subtract the spatial derivative of \Eq{dnu_dt} from the temporal derivative of \Eq{dn_dt}; then substitute $\partial_x E = - 4 \pi e n$, which flows from \Eq{dE_dx}, and integrate over $x$. Assuming that $n$ is small enough, we neglect terms nonlinear in $E = O(n)$, and we also neglect the slow driving force, $O(E^0 A^1)$, as nonresonant to the rapid plasma oscillations. As we estimate below, resonant terms of higher orders in $A$ can be dropped too. To the lowest order in $\omega_d$ \cite{comment1}, this gives the following linear dimensionless equation for $E$,
\begin{gather} \label{d2E_dt2}
 \frac{\partial^2 E}{\partial t^2}
 - 3\,\frac{\partial^2 E}{\partial x^2}
 + E  = - \widetilde{n}_d E 
 + \widetilde{R}\,\frac{\partial}{\partial x}\left(\widetilde{n}_d\,\frac{\partial E}{\partial x}\right).
\end{gather}
Here and further we measure time in units $\omega_p^{-1}$ and length in units $v_{th}/\omega_p$; also, $\widetilde{n}_d = n_d/n_0$, and $\widetilde{R}=R n_0/(m v_{th}^2)$. 
Next, we decompose the field into unperturbed eigenmodes, 
\begin{gather}
 E = \text{Re} \sum_{m=1}^{\infty} E_m e^{-i \omega_m t} \sqrt{2/\ell}\, \sin(k_m x),
\end{gather}
where $E_m(t)$ are complex coefficients. To zeroth order in $A$, \Eq{d2E_dt2} yields the dimensionless dispersion relations $\omega_m^2 = 1 + \beta m^2$, where $\beta = 3\pi^2/\ell^2$ is analogous to the anharmonicity parameter in a quantum oscillator \cite{ido5}. We also assume $\beta m^2 \ll 1$, as usual. It is convenient to introduce new variables $\psi_m$ via $\rho_m \psi_m(t)=E_m e^{-i\omega_m t}$. Here $\rho_m$ are constants such that in the unperturbed system, $|\psi_m|^2$ are the actions of individual modes; specifically one finds $\rho_m = (8\pi/\omega_m)^{1/2}$ \cite{ilya_PoP09}. The equations for $\psi_m$, obtained via Fourier-transforming \Eq{d2E_dt2}, are:
\begin{gather}
i \dot{\psi}_m=\omega_m \psi_m + \sum_{m'}h_{m,m'}\psi_{m'}, \label{idEn_dt}\\ 
h_{m,m'}=\frac{\rho_m \rho_{m'}}{8\pi\ell} \int_0^{\ell} \sin(k_m x) \mathcal{\hat{F}}\sin(k_{m'} x)\,dx, \label{coupling}  
\end{gather}  
where $\mathcal{\hat{F}}=\widetilde{n}_d - \partial_x\,\widetilde{R}\,\widetilde{n}_d\,\partial_x$ is a differential operator. Since $h_{m,m'}$ is Hermitian, the evolution of $\psi_m$ is manifestly unitary;  \ie the wave total action $\sum_m |\psi_m|^2$ is conserved. The vector $\psi = (\psi_1, \psi_2, \dots)$ can then be understood as the plasmon wave function in the energy representation. Next, we will assume the resonance condition $\omega_d\approx \omega_{m,m + N}$, where the level spacing is $\omega_{m, m+N} = \omega_{m+N}-\omega_{m} \approx \beta N (m + N/2)$. Then, $\beta \sim \omega_d \ll 1$, and therefore, in the already small coupling term, we must adopt  $\omega_m \approx 1$, \ie  $\rho_m \approx \sqrt{8\pi}$ and neglect the higher order terms in pressure. This leads to
\begin{gather}
i \dot{\psi}_m = \omega_m \psi_m + \frac{A}{4} \left(\psi_{m - N} + \psi_{m + N} \right) \cos \varphi_d.
\label{psi_n}
\end{gather}
Note also that, for $\beta m^2 \ll 1$, \Eq{psi_n} is equivalent to the energy representation of the Schr\"odinger equation for a quantum particle in a square potential well \cite{comment3}. This can be understood as  follows: at weak spatial dispersion, all $\omega_m$ are close to $\omega_p$, so \Eq{d2E_dt2} permits a quasioptical approximation,  turning into the standard quantum Schr\"{o}dinger equation with $\widetilde{n}_d$ serving as an effective potential. 
(This analogy has been noted, \eg in Ref.~\cite{Dewar72}.)

\begin{figure}[tb]
\includegraphics[width=8.5cm]{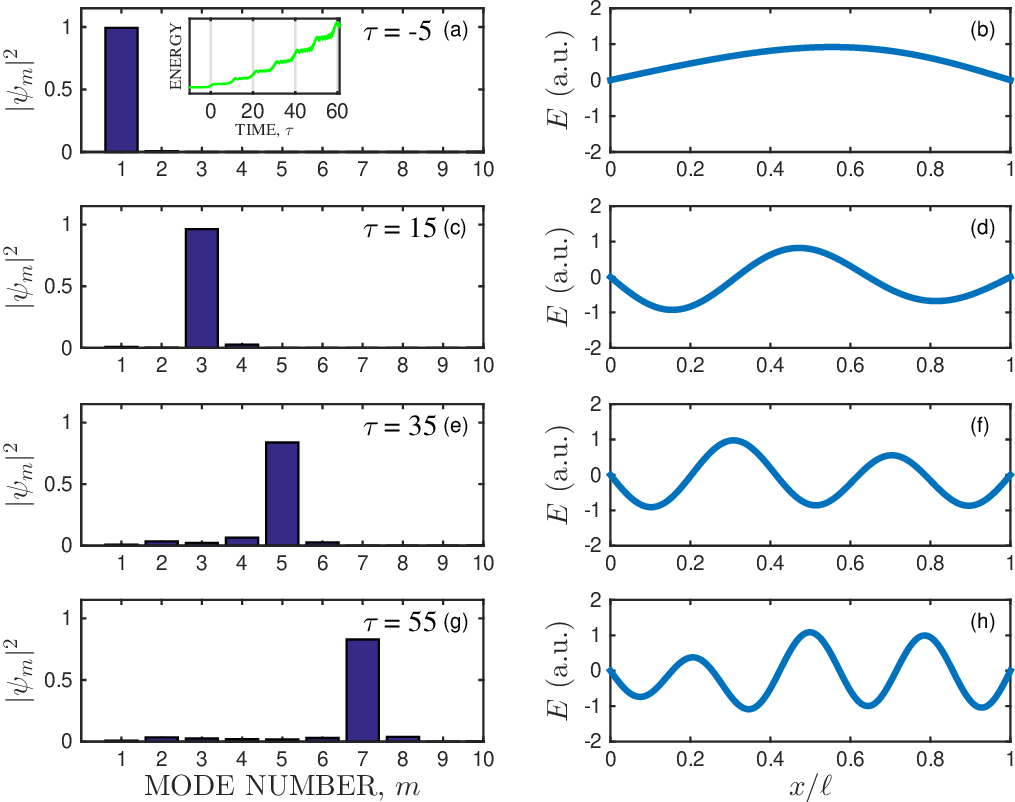}
\caption{(color online) Langmuir wave evolution in the LC regime ($P_2=10$). Left -- spectral representation; snapshots of the occupation numbers, $|\psi_m|^2$, vs mode numbers, $m$ at times $\tau=[-5, 15, 35, 55]$. Right -- spatial representation; snapshots of the wave field, $E$, vs $x$ at the same times. The inset in subplot (a) shows the energy growing with time in a ladder-like manner, where the transitions occur at the theoretically predicted times $\tau_m = m P_2$.}
\label{fig1}
\end{figure}

\msection{LC regime} Classical AR was previously studied in the infinite square potential well as a limiting case $(j\to \infty)$ of the potential $V_0 = x^{2j}$ \cite{lazar87}, but quantum LC was not \cite{comment2}. For studying  LC in our system, we consider the modulation \eq{n_d} with $N = 1$ and a monotonically increasing driving frequency, $\omega_d = \omega_{1,2} + \alpha t$, where $\alpha > 0$. At $t = 0$, this will initiate resonant transitions between levels $m = 1$ and $m = 2$, which we denote as $1 \to 2$. Later, transitions between higher levels, $m \to m + 1$, occur when the resonance condition, $\omega_d = \omega_{m, m + 1}$, is satisfied. Following the quantum LC theory \cite{Marcus_th,ido5}, we define slow time $\tau=\sqrt{\alpha}t$, driving parameter $P_1 = A/(4\sqrt{\alpha})$, and anharmonicity parameter, $P_2=\beta/\sqrt{\alpha} $, which plays the role of an effective Planck constant in the classical system. If $P_2 \gg 1 + P_1$, the system remains in the quantum LC regime, when only two levels are resonantly coupled at any given time. Then, transitions $m \to m + 1$ occur at times $\tau_m = m P_2$,
where each such transition can be described by the commonly known LZ theory \cite{LZ,Tracy} and thus has a probability 
\begin{gather}
\mathcal{P}_{m\to m+1}=1-\exp\left(-\pi P_1^2 /2\right). \label{LZ}
\end{gather}
At large enough $P_1$, $\mathcal{P}_{m \to m + 1} \approx 1$, \ie all quanta are being transferred, and the dynamics in this limit is characterized by successive two-level LZ transitions.

\begin{figure}[t]
\includegraphics[width=8.5cm]{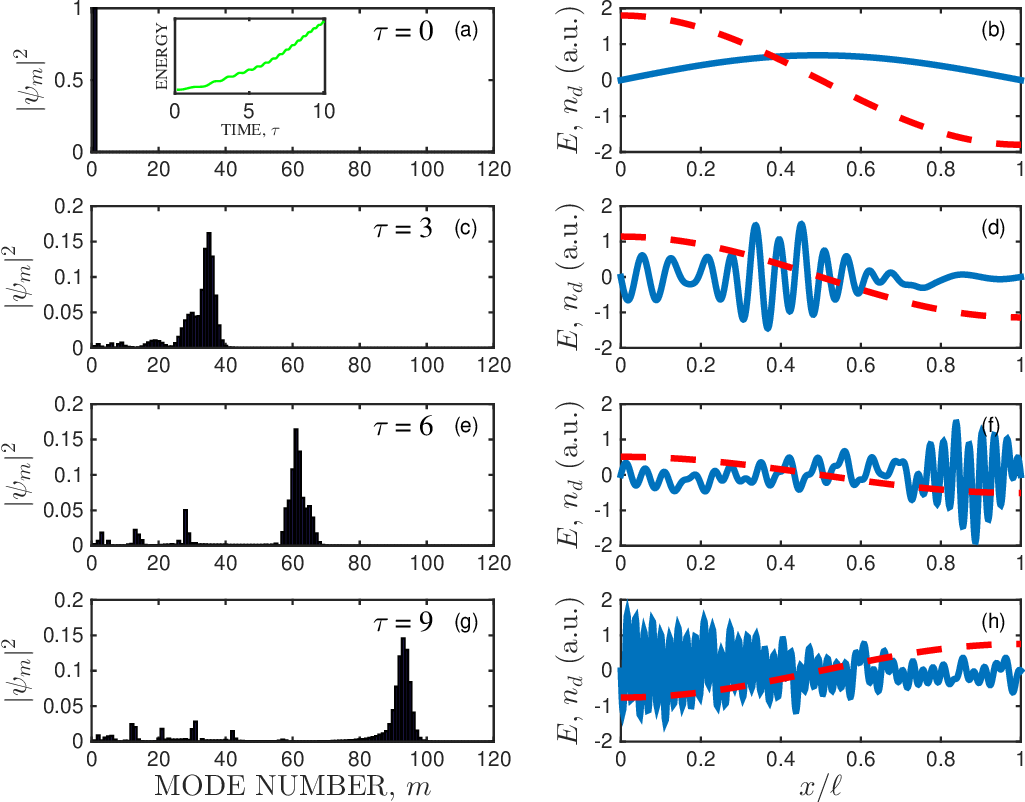}
\caption{(color online) Langmuir wave evolution in the AR regime ($P_2 = 0.1$).	Left -- spectral representation; snapshots of the occupation numbers, $|\psi_m|^2$, are shown vs mode numbers, $m$, at times $\tau = [0, 3, 6, 9]$, illustrating an autoresonant plasmon trapped in a density modulation effective potential. Right -- spatial representation; snapshots of the wave field, $E$ (solid blue lines), and the driving density modulations, $n_d$ (dashed red lines), are shown vs $x$ at the same times. The inset in subplot (a) shows the energy growing with time.}
\label{fig2}
\end{figure}

In \Fig{fig1}, we illustrate this LC dynamics of Langmuir modes by numerically simulating \Eq{psi_n} with $N = 1$ and initial conditions $\psi_m(t = t_0) = \delta_{m,1}$. We choose $\{\alpha, \beta, A\} =\{10^{-8}, 10^{-3}, 6 \times10^{-4}\}$, so $P_2 = 10$, and $P_1 = 1.5$. In the figure, snapshots of the levels population (a,c,e,g) and the electric field (b,d,f,h) are shown for multiple $\tau$, respectively. The inset also shows the total energy, $\sum_m \omega_m |\psi_m|^2$,
which is seen to increase with time in a ladder-like manner. The moments of time at which $m \to m + 1$ transitions occur agree with the theory, which predicts $\tau_m = m P_2$ \cite{ido5}. The $m$th-level occupation numbers, $|\psi_m|^2$, also agree with the theoretical predictions, namely, $|\psi_m|^2 = \prod_{j = 1}^{m - 1} \mathcal{P}_{j\to j+1}$, where $\mathcal{P}_{j \to j+1}$ are given by \Eq{LZ}. For example, at the adopted parameters, one has $|\psi_7(\tau = 55)|^2 = 0.83$, which deviates by only about 1\% from the value obtained in the simulation.

A similar argument justifies our neglecting the terms of higher order in $A$ in the derivation of \Eq{d2E_dt2}. A direct calculation shows that those terms have the form $O(\omega_d E^1 A^2)$, so they cause additional ``subharmonic'' resonant transitions $m \to m + 2$ with the effective driving parameter $\widetilde{P}_1 \sim \beta A^2/\sqrt{\alpha}$. The corresponding probability $\widetilde{\mathcal{P}}_{m \to m + 2}$  is given by an expression similar to \Eq{LZ}, but $P_1$ should be replaced with $\widetilde{P}_1$ \cite{ido7}. For small enough $A$, one obtains $\widetilde{\mathcal{P}}_{m \to m + 2} \sim \beta^2 A^4/\alpha \ll 1$. To have both $\mathcal{P}_{m \to m + 1} \approx 1$ and $\widetilde{\mathcal{P}}_{m \to m + 2} \ll 1$, one must then require the condition (which is satisfied in our simulations)
\begin{gather}\label{eq:A}
\alpha^{1/2} \lesssim A \ll \beta^{-1/2} \alpha^{1/4}.
\end{gather}

Also note another, kinetic restriction of the LC mechanism. It stems from collisionless dissipation, which is not contained in our fluid equations. The local Landau damping rate for the $m$th mode in Maxwellian plasma is given by $\gamma_m \approx k_m^{-3} \sqrt{\pi/8}\,\exp(-k_m^{-2}/2-3/2)$ \cite{physical_kinetics}, where $k_m = m\sqrt{\beta/3}$, as before. Therefore, during the transition between neighboring levels, which occurs on the time scale $\Delta t = (\tau_{m+1} - \tau_m)/\sqrt{\alpha} = \beta/\alpha$, the energy decreases by the factor $\exp(2 \Gamma_m)$, $\Gamma_m = \gamma_m \Delta t$. The value of $\Gamma_m$ grows rapidly with $m$. Hence, just as the wave energy is shifted to the mode with $m = m_{\rm kin}$, defined as that having $\Gamma_m \sim 1$, the energy is transferred to electrons almost momentarily, heating the tail distribution much like in \Ref{schmit}. At the parameters used in \Fig{fig1}, $m_{\rm kin} = 9$. This is larger than the maximum $m$ attained in the simulation, so neglecting Landau damping is justified.

\begin{figure}[t]
\includegraphics[width=8.5cm]{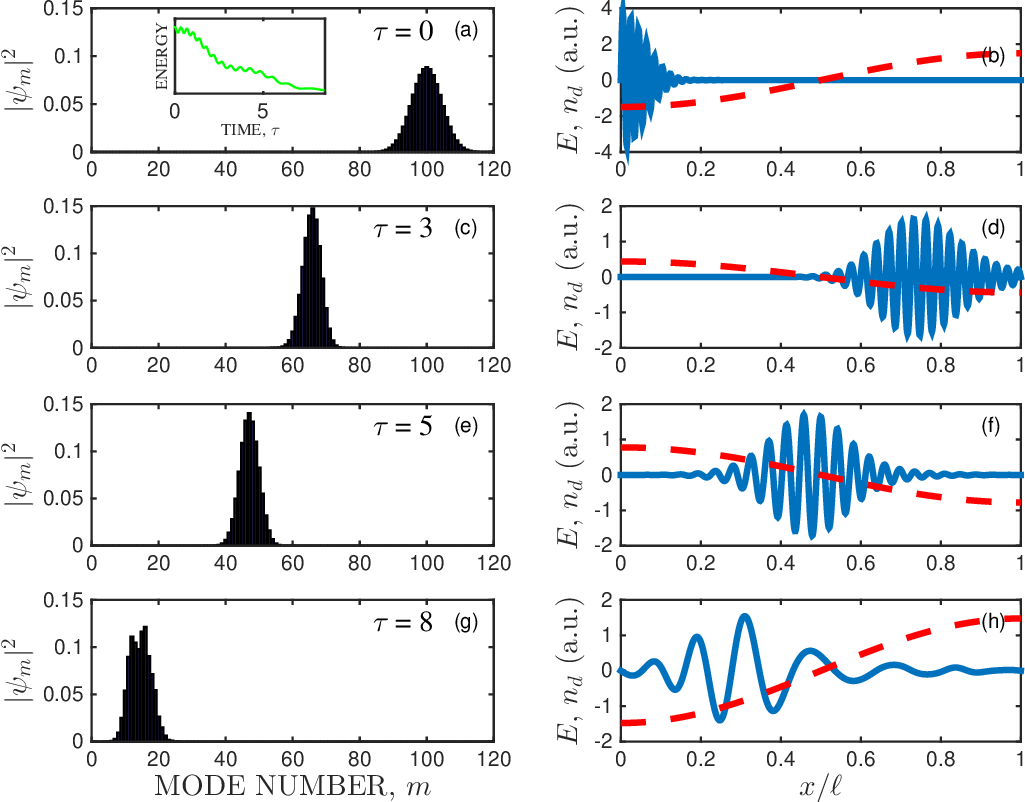}
\caption{(color online) Langmuir wave evolution in the down-chirped AR regime ($P_2 = 0.1$). The plotted quantities are the same as in \Fig{fig2}, but the wave is initialized at larger $m$, and the chirp of the modulation frequency has the opposite sign.}
\label{fig3}
\end{figure}

\msection{AR regime} In contrast to the LC dynamics, in the limit $P_2 \ll 1$, many levels are coupled simultaneously. It can be shown then (\eg  using the Wigner phase space approach, as in Ref. \cite{ido5}) that quantum LC continuously transforms into the classical AR by decreasing the effective Planck constant, $P_2$.
The electric field dynamics is then understood as AR acceleration of plasmons that satisfy $\omega_d = k_d v_g$, where 
\begin{gather}
v_g = \frac{\omega_{m + 1}-\omega_{m}}{k_{m + 1} - k_m} \approx \frac{\partial \omega}{\partial k}.
\end{gather}
Notably, this is the same ``group-resonance'' condition that was recently discussed in \Ref{ilya_PRL14} (see \Ref{Tsytovich} too). Also notably, the AR acceleration of plasmons that we report here is akin to AR acceleration of resonant electrons in phase-mixed nonlinear waves such as Bernstein-Greene-Kruskal waves \cite{BGKapp,lazar111_ido2,Breizman_Schmit}. The AR dynamics of Langmuir waves is illustrated in \Fig{fig2}, for which we adopted $\{\alpha,\beta,A\} =\{10^{-8}, 10^{-5}, 4.8 \times10^{-3} \}$, so $P_2 = 0.1$. The simultaneous coupling of many levels is clearly seen in the left subplots, while the right subplots present the electric field of the autoresonant plasmon (solid blue lines) driven by the chirped density modulations, $\widetilde{n}_d$ (red dashed lines). In this case, Landau damping affects the total wave energy only by about $1\%$ overall, so the kinetic restriction is not essential.
 
Interestingly, the dynamics is reversible, at least approximately. In fact, one can just as well capture a wave envelope at large $k$ and then transport it down the spectrum, much like a trapped charged particle can be decelerated by a resonant field. We demonstrate the effect in \Fig{fig3}, where the initial conditions are
$E_0(x)\sim\exp(-(x-x_0)^2/2\sigma^2) \sin\left(k_{m_r}x \right)$, where $m_r$ is the resonant wave number. In this example, we apply down-chirped driving phase
$\varphi_d=\omega_{m_r,m_r+1} t-\alpha t^2/2+\varphi_0$ with $\{\beta, x_0, \sigma, m_r, \alpha, A, \varphi_0\} = \{10^{-5},0,0.05,100, 10^{-8}, 0.004, \pi\}$, where
 $\varphi_0$ was chosen such that the plasmon is initially phase-locked with $n_d$.

\msection{Variational formulation} Let us now recast our theory in a form that is not restricted to Langmuir waves but allows extending the above results to general nondissipative linear waves. Any such wave, described by some real field $\vec{E}(t, \vec{x})$ (electric field being an example), can be assigned a Lagrangian bilinear in $\vec{E}$, namely, of the form~\cite{ilya_PLA14}
\begin{gather}
L = \int \vec{E} \cdot \oper{\mc{D}}(t, \vec{x}, i\pd_t, -i\del) \cdot \vec{E} \,d^3x.
\label{Lgen}
\end{gather}
The differential operator $\oper{\mc{D}}$ can be considered Hermitian without loss of generality. Suppose now that $\oper{\mc{D}} = \oper{\mc{D}}_0 + \oper{\mc{D}}_d$, where $\oper{\mc{D}}_0$ is some Hermitian operator that determines the Lagrangian in stationary homogeneous medium, and $\oper{\mc{D}}_d$ is a Hermitian operator that governs the mode interaction driven by some weak modulation. Let us represent the field as $\vec{E}(t, \vec{x}) = \text{Re}\, \sum_m \mc{E}_m(t)\,\vec{e}_m(\vec{x})$. Here $\mc{E}_m$ are complex amplitudes and $\vec{e}_m$ are orthonormal eigenmodes corresponding to the eigenfrequencies $\omega_m$ of $\oper{\mc{D}}_0$. Then, $L = \sum_m L_m + L_{d}$, where the two terms are, respectively, due to $\oper{\mc{D}}_0$ and $\oper{\mc{D}}_d$. Specifically~\cite{ilya_PLA14},
\begin{gather}
L_m = \frac{i}{2}\,(\psi^*_m \dot{\psi}_m - \dot{\psi}^*_m \psi_m) - \omega_m|\psi_m|^2,
\end{gather}
where the complex amplitudes $\psi_m$ are defined such that $|\psi_m|^2$ is the action of the $m$th unperturbed mode; \ie $\rho_m \psi_m = \mc{E}_m$, where $\rho_m^{-2} = d \bar{L}_m/d\omega_m$, $\bar{L}_m = (1/2) \int \vec{e}_m^* \cdot \oper{\mc{D}}_0 \cdot \vec{e}_m \,d^3x$, and $\omega_m$ are found by solving $\bar{L}_m(\omega_m) = 0$ \cite{ilya_PRA12}. Also, it is easy to see that 
\begin{gather}
L_{d} \approx \sum_{m,m'} \psi_m^* h_{m,m'} \psi_{m'}, \\ 
h_{m,m'} =\frac{1}{2}\,\rho_m \rho_{m'} \int \vec{e}_m^* \cdot \oper{\mc{D}}_{d} \cdot 
\vec{e}_m \,d^3x.
\end{gather}
Then the equation for $\psi_m$ has the form \eq{idEn_dt}, and, since $h_{m,m'}$ is Hermitian, it manifestly conserves the wave total action, $\sum_m |\psi_m|^2$. For electromagnetic waves in particular, one has $\oper{\mc{D}}_d = \oper{\chi}_d/(8\pi)$, where $\oper{\chi}_d$ is the modulation-driven perturbation to the medium susceptibility. This can be seen, for instance, by comparing \Eq{Lgen} with its geometrical-optics limit \cite{ilya_PRA12}. In the case of Langmuir waves, $\oper{\chi}_d = \mathcal{\hat{F}}$, and $\oper{\mc{D}}_0 = \oper{\epsilon}/(8\pi)$, where $\oper{\epsilon}$ is the dielectric permittivity operator; then one recovers \Eq{psi_n}. Finally, we note that it is straightforward to apply this approach to other classical waves \cite{Winn}; hence, our further observations  of LC and AR apply too. 

\msection{Summary} We report the first theoretical prediction of LC in a classical system. Specifically, we show that quasiperiodic chirped modulations of the background density can couple discrete eigenmodes of bounded plasma to produce controllable shift of the wave spectral energy distribution. Apart from academic interest, the new method of continuously controlling the wavelength of Langmuir oscillations might find practical applications, such as regulating coherent Raman scattering of laser radiation in plasma for generating short ultraintense pulses \cite{nat_BRA}. Our results indicate how similar techniques might be practiced with other plasma modes too, or in other settings such as waveguides or photonic crystals. Our work also bridges a number of effects that were previously considered unrelated.  In particular, plasmon acceleration reported here can be seen as the resonant counterpart of the adiabatic ponderomotive effects on waves discussed recently in Ref. \cite{ilya_PRL14}. It is also akin to the resonant acceleration of charged particles trapped by chirped nonlinear plasma waves \cite{BGKapp,Breizman_Schmit}. Finally, our results further advance the general idea \cite{Dragoman,Tracy,ilya_PLA14}  that posing classical waves in quantumlike terms can be quite fruitful.

The work was supported by NNSA grant DE274-FG52-08NA28553, DOE contract DE-AC02-09CH11466, and DTRA grant HDTRA1-11-1-0037.


\begin{thebibliography}{99}
\bibitem{comment1} 
The omitted term is of the form $O(\omega_d E)$. Under the assumed conditions, keeping it would modify the mode-coupling coefficients below by factor $1 + O(\omega_d)$. This correction is negligible within the adopted accuracy, because we further assume $\omega_d = O(\beta)$ and entirely neglect thermal corrections in \Eq{psi_n}.
\bibitem{comment3} 
Our system differs from the anharmonic quantum oscillator \cite{ido5} just by the values of $\omega_m$ and 
$\rho_m$.
\bibitem{comment2}
In an infinitely deep square potential, a quantum particle  (or, equivalently, classical wave)  can be captured into AR even at the lowest energy state, since the corresponding eigenfrequency is nonzero. This is in contrast with the classical model of AR in such systems, within which automatic capturing into AR at the very bottom of an infinitely deep square potential is impossible \cite{lazar87}.
\end{thebibliography}
\end{document}